\documentclass{edp-conf}
\usepackage{graphicx}
\begin{document}

\TitreGlobal{SF2A 2005}

%%-----------------------------
%%      the top matter
%%-----------------------------
\title{The puzzle of the soft X-ray excess in AGN: absorption or reflection?}
\author{Chevallier, L.}\address{LUTH, Observatoire de Paris, Section de Meudon, F-92195 Meudon Cedex, France;
\email{loic.chevallier@obspm.fr\ \&\ suzy.collin@obspm.fr\ \&\ anne-marie.dumont@obspm.fr\\ \&\ anabela.goncalves@obspm.fr\ \&\ rene.goosmann@obspm.fr}}
\author{Collin, S.}\sameaddress{1}
\author{Dumont, A. M.}\sameaddress{1}
\author{Czerny, B.}\address{Copernicus Astronomical Center, Bartycka 18, 00-716 Warsaw, Poland;\\ \email{bcz@camk.edu.pl}}
\author{Mouchet, M.}\address{APC, Universit\'e Denis Diderot, F-75005 Paris, France; \email{martine.mouchet@obspm.fr}}
\author{Goncalves, A. C.}\sameaddress{1}\secondaddress{CAAUL, Observat\'orio Astron\'omico de Lisboa, Tapada da Ajuda, 1349-018 Lisboa, Portugal}
\author{Goosmann, R.}\sameaddress{1}
\runningtitle{Chevallier et al.: The puzzle of the soft X-ray excess in AGN}
\setcounter{page}{1}
% Keep this line, even if the page will be settled afterwards..
%\author{Chevallier, L.}
%\author{Collin, S.}
%\author{Czerny, B.}
%\author{Mouchet, M.}
%\author{Gon\c{c}alves, A. C.}
%\author{Goosmann, R.}
% Repeat the authors here, this will help to make the final index

\maketitle
\begin{abstract}
The 2-10 keV continuum of AGN is generally well represented by a single power law. However, at smaller energies the continuum displays an excess with respect to the extrapolation of this power law, called the ``soft X-ray excess''. Until now this soft X-ray excess was attributed, either to reflection of the hard X-ray source by the accretion disk, or to the presence of an additional comptonizing medium, giving a steep spectrum. An alternative solution proposed by Gierli\' nski and Done (2004) is that a single power law well represents both the soft and the hard X-ray emission and the impression of the soft X-ray excess is due to absorption of a primary power law by a relativistic wind. We examine the advantages and drawbacks of reflection versus absorption models, and we conclude that the observed spectra can be well modeled, either by absorption (for a strong excess), or by reflection (for a weak excess). However the physical conditions required by the absorption models do not seem very realistic: we would prefer an ``hybrid model''.
\end{abstract}
%
%%-----------------------------
%%      your text
%%-----------------------------

\begin{figure}[t]
 \centering
 \includegraphics[width=7.2cm]{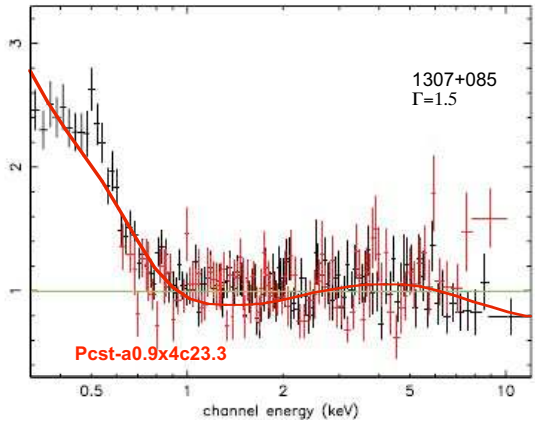}
 \caption{Comparison between the observed X-ray spectra for PG 1307+085 (Piconcelli et al.~2005, Fig.~2) and the computed spectrum (arbitrary units) of a power law primary continuum $\alpha=0.9$ absorbed by a constant total pressure slab ($\xi=10^4$, $N =  2\ 10^{23}$, $v/c = 0.2$, CGS units). The observed and computed spectra have been both divided by the observed power law of photon index $\Gamma=1.5$ over the 2--10 keV range.
}
 \label{fig-comp-mod-obs}
\end{figure}

\section{Introduction}

Now more than 50\% of well studied Seyfert 1 galaxies and many quasars are known to possess absorbers (e.g., Blustin et al. 2005).
One main issue is to explain the apparent change of slope in the overall X-ray spectrum at $\sim$ 1 keV in Sy1s and QSOs.
When fitting an observed X-ray spectrum with a power law plus absorption plus (eventually) the Compton reflection component plus (eventually) the iron line and (eventually) narrow spectral features, the model usually underpredicts the observed spectrum in the soft X-ray range. An additional component -- a soft X-ray excess -- is needed (Wilkes \& Elvis 1987).
Apart from an usual additional continuum or a strongly ionized reflection, this component is well modeled by absorption of an originally rather soft power law intrinsic spectrum due to an absorber having a random or bulk velocity of several thousands of km~s$^{-1}$ (Gierli\' nski \& Done 2004).
We consider the advantages and drawbacks of the reflection vs. absorption models (Chevallier et al.~2005), using our code TITAN (Dumont et al.~2003).

\section{Results}

Gierli\' nski \& Done (2004) modeled the X-ray spectrum of the Narrow Line Seyfert 1 PG 1211+143 by a steep power law continuum between 0.1 and 20 keV, absorbed by an ionized slab of constant density.
Note that the observed spectrum is smeared by a large (Gaussian) velocity dispersion $v/c= 0.2$ in order to get  a ``quasi-continuum'' with no narrow features. This high velocity can be due to an accelerated outflow, or to a disk wind dominated by Keplerian motion and produced very close to the black hole.

We have computed the absorption spectra -- no corresponding emission and the absorbing medium covers completely the primary source of radiation -- for a grid of constant density models.
Any small variation of the parameters, like the column-density $N$, the ionization parameter $\xi=L/nR^2$ ($L$ is the luminosity of the primary source, $n$ and $R$ are the hydrogen number density and the distance from the primary source, respectively, of the medium at the illuminated surface), or the slope $\alpha$ of the primary continuum, would induce a strong variation on the shape of the X-ray spectrum.

\begin{figure}[t]
 \centering
 \includegraphics[width=6.22cm]{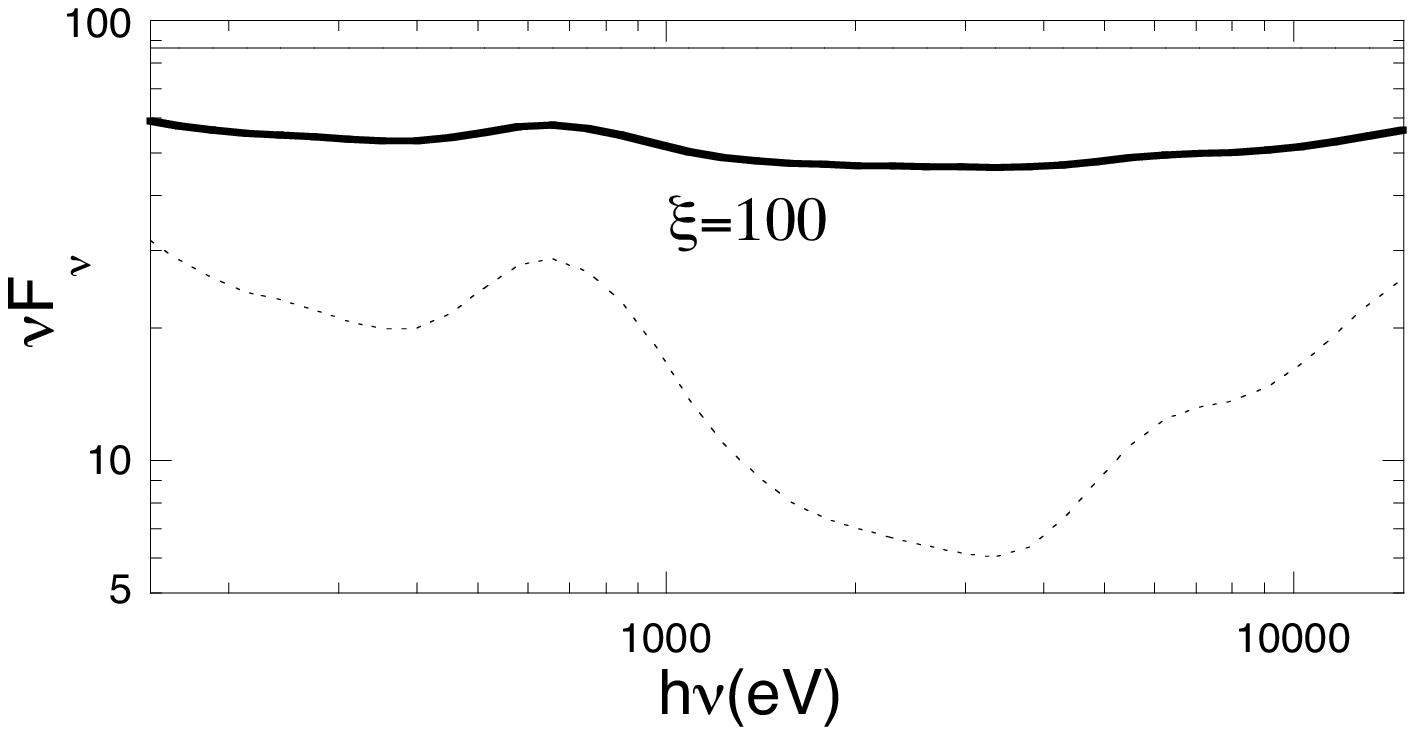}
 \includegraphics[width=6cm]{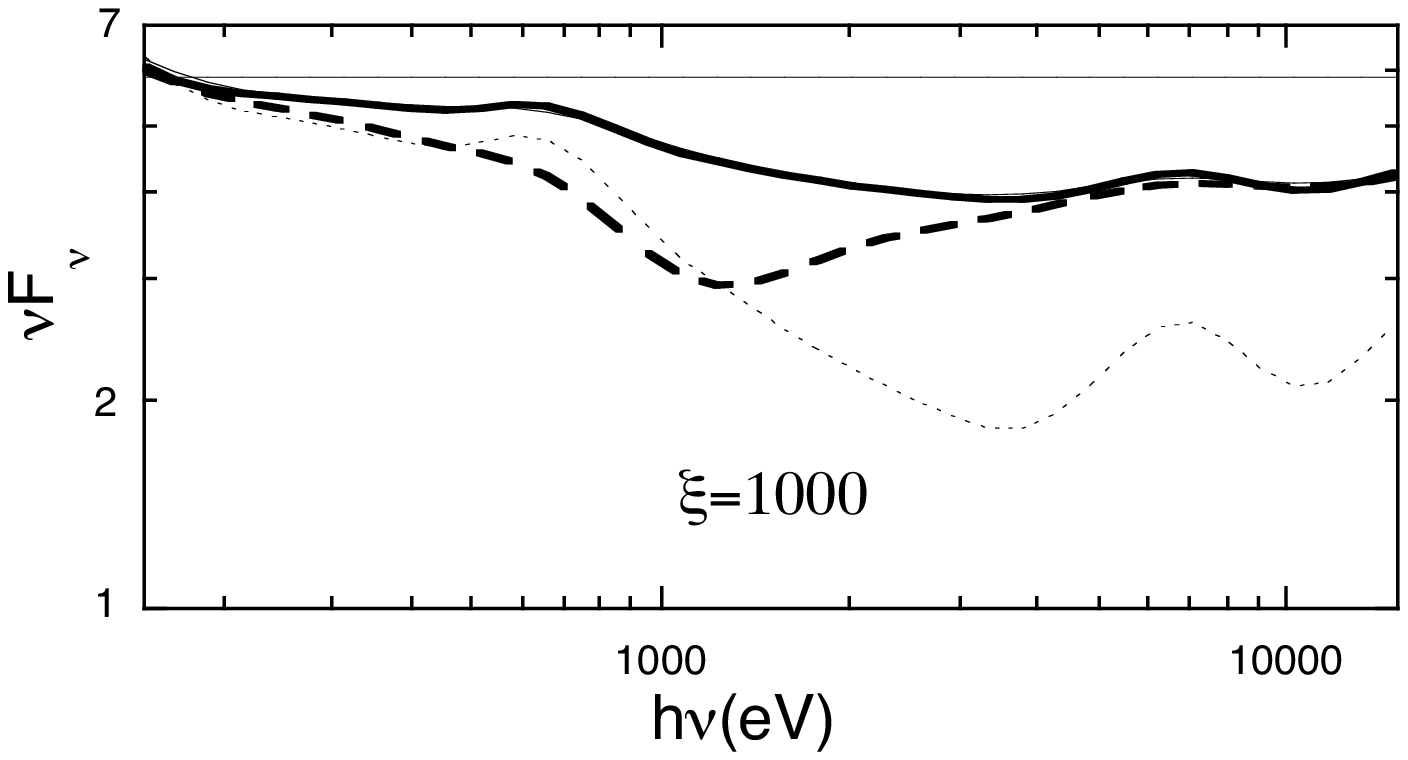}
\caption{Computed spectra (arbitrary units) for thick reflection models (left: $\xi$=100, right: $\xi$=1000). The components shown are the primary continuum (straight lines), the reflection spectrum (thin dots), half sum of reflection plus primary continuum (thick lines) which may be directly ``observed'' or absorbed by a constant total pressure slab $N=10^{22}$, $\xi=100$ (CGS units), with a dispersion velocity $v/c=0.2$ (thick dashes).}
 \label{fig-x2x3-spe}
\end{figure}

Such a variation is not observed from one object to the other.
It is more appropriate to assume that the absorbing medium is in total -- gas and radiation -- pressure equilibrium owing to the short dynamical time scale needed to reach again an equilibrium (less than one day for $R \sim 10\ R_{\rm G}$, where $R_{\rm G}=GM/c^2$, and $M \sim 10^7\ M_\odot$).
The thickness of the slab cannot then exceed a maximum value for a given $\xi$, due to thermal instabilities. A consequence is the existence of a ``maximum absorption trough", which cannot be exceeded.

Figure \ref {fig-comp-mod-obs} shows as an exemple the comparison of the X-ray spectrum of PG 1307+085 (Piconcelli et al.~2005) with that obtained with an absorbing slab of constant total pressure.
This spectrum is  well fitted, considering that the narrow emission feature around 0.5 keV -- the OVII complex -- must be provided by another emitting region.

Absorption models are not very satisfactory from a physical point of view.
Owing to its large column density ($\sim 3\ 10^{23}$ cm$^{-2}$), the wind implies too massive outflows (near the Eddington limit).
Both the wind and the accretion models require an additional UV emission, which has to be provided by a geometrically thin accretion disk.
The coexistence of a spherical accretion flow at about $25\ R_{\rm G}$ and a thin disk seems quite artificial.
So we come back now to the ``traditional'' reflection models, including a ``cold'' accretion disk emitting the UV, surrounded by a hot corona emitting X-rays which are reflected by the disk (Haardt \& Maraschi 1993).
Figure \ref{fig-x2x3-spe} shows two examples of reflection models, with $\xi=100$ and 1000.
The soft X-ray excess (thick line) almost disappears when the reflection spectrum (thin dots) is added to the primary continuum (thin line). Either $\xi$ is small and the reflection spectrum displays a strong X-ray excess, but it is negligible as compared to the primary one; or $\xi$ is large, and the reflection spectrum is comparable in flux to the primary one, but it has a small X-ray excess. A strong excess requires to hide the primary continuum (Fabian et al. 2002, Crummy et al. 2005).

Since both the absorption and the reflection models seem inadequate, we propose an ``hybrid model", including the traditional reflection model, plus a high velocity absorbing medium with a modest thickness.
As shown by Fig.~\ref{fig-x2x3-spe} (right panel), the small soft X-ray excess displayed by the reflection model (thick line) is increased when it is absorbed by such a wind (thick dashes) and becomes comparable to the observations.
The mass outflow rate is thus 30 times smaller than in the previous absorption model.

\section{Conclusion}

Absorption models could account for some strong soft X-ray excesses, but require a kind of ``fine tuning" in order to constrain the 1 keV trough, which otherwise could have any strength (e.g., constant density models). We have suggested a medium in total -- gas and radiation -- pressure equilibrium, which leads to a maximum intensity of the trough, as well as a ``universal" shape of this maximum trough, due to the thermal instability mechanism. A complete grid of constant total pressure models, very demanding in computation time, is necessary to pursue this study.
In the absorption model, either a thick accretion flow, or a relativistic wind is required. None of them seem realistic from a physical point of view, and moreover both require an additional source of UV emission, like a geometrically thin accretion disk.
On the other hand, the traditional reflection models involving a hot corona emitting X-rays which are reflected by a cold disk cannot account for the observations, unless the X-ray source is hidden from our view.
Therefore we favor an ``hybrid" model, where the primary UV-X source could be produced by a disk-corona system, and then absorbed by a modest relativistic wind.

\begin{acknowledgements}
A.\,C. Gon\c{c}alves acknowledges support from the {\it {Funda\c{c}\~ao para a Ci\^encia e a Tecnologia}}, Portugal, under grant no. BPD/11641/2002. Part of this work was supported by the Laboratoire Europ\'een Associ\' e Astrophysique Pologne-France.
\end{acknowledgements}

%%-----------------------------
%%      your bibliography
%%-----------------------------
%In preparing the reference list please adhere to the following format.
% Attention should be paid to the order of the items in each reference
% and to the punctuation used. Please see Sect. 4 in the User's Guide
% that comes with the new macro package.

\end{document}